\documentclass[aps,reprint,superscriptaddress,twocolumn,longbibliography]{revtex4-2}
\usepackage[english]{babel}
\usepackage[utf8]{inputenc}
\usepackage[colorinlistoftodos, color=green!40, prependcaption]{todonotes}
\usepackage{amsthm}
\usepackage{amsfonts}
\usepackage{mathtools}
\usepackage{physics}
\usepackage{xcolor}
\usepackage{graphicx}
\usepackage[left=23mm,right=13mm,top=35mm,columnsep=15pt]{geometry} 
\usepackage{adjustbox}
\usepackage{placeins}
\usepackage[T1]{fontenc}
\usepackage{lipsum}
\usepackage{csquotes}
\usepackage[pdftex, pdftitle={Article}, pdfauthor={Author}]{hyperref} 
\usepackage{caption}
\usepackage{subcaption}
\usepackage{bm}

\usepackage{ragged2e}
\makeatletter
\renewcommand{\@makecaption}[2]{%
  \vskip\abovecaptionskip
  \justifying
  \small #1: #2\par
  \vskip\belowcaptionskip}
\makeatother

\begin{document}

\preprint{APS/123-QED}

\title{Ghost-RISB for Correlated Electron-Phonon Systems: Application to the Hubbard-Holstein Model}

\author{Samuele Giuli}
    \email[Correspondence email address: ]{sgiuli@flatironinstitute.org}
    \affiliation{Center for Computational Quantum Physics, Flatiron Institute, 162 5th Avenue, New York, New York 10010, USA}

\author{Ricardo J. Campos-Lopes}
    \affiliation{International School for Advanced Studies (SISSA), via Bonomea 265, 34136 Trieste, Italy}

\author{Emin Moghadas}
    \affiliation{Institute of Solid State Physics, TU Wien, 1040 Vienna, Austria}
    \affiliation{Center for Computational Quantum Physics, Flatiron Institute, 162 5th Avenue, New York, New York 10010, USA}

\author{Massimo Capone}
    \affiliation{International School for Advanced Studies (SISSA), via Bonomea 265, 34136 Trieste, Italy}

\date{\today} 

\begin{abstract}
We develop a generalization of the ghost-rotationally-invariant slave-boson (ghost-RISB) method that incorporates local phonon modes coupled to arbitrary on-site electronic degrees of freedom, enabling a nonperturbative treatment of electron-electron and electron-phonon interactions within an efficient variational framework. The method extends the ghost-orbital construction to capture dynamical self-energy effects and phonon-induced renormalizations beyond static slave-boson approaches.
Benchmarking against dynamical mean-field theory (DMFT) results for the Hubbard-Holstein model, we find excellent quantitative agreement for electron quasiparticle weights and phonon properties across a wide range of coupling strengths, and it captures accurately the competition between electron-electron and electron-phonon interactions. We show that the inclusion of the \textit{ghost} orbitals is crucial to accurately describe the regime of low-frequency, strongly dynamical, phonons. The extended ghost-RISB achieves this accuracy at a fraction of the computational cost of DMFT, due to its self-consistency rooted in static observables instead of dynamical ones, enabling rapid exploration of correlated electron-phonon phase diagrams. We exploit this advantage to characterize the most demanding regime of strong coupling and adiabatic phonons. Our analysis shows a suppression of the superconducting order parameter, which is interpreted as a Franck-Condon-like reduction of the overlap between the phonon wavefunctions associated with empty and doubly-occupied sites in the bipolaronic regime.
\end{abstract}

\keywords{Hubbard model, Holstein model}

\maketitle

\section{Introduction} \label{sec:introduction}

Electrons in solids can display remarkable phenomena resulting from their mutual interaction (electron-electron correlation) and from the coupling with lattice vibration (electron-phonon coupling). Just to mention a few, electron-electron repulsion can drive Mott physics, emergent quasiparticles, and nontrivial multiplet effects in multiorbital systems and -most likely- unconventional superconductivity.  Electron-phonon (e-ph) interactions can induce polaron formation, charge ordering, structural transitions, or conventional superconductivity. Yet, advances in synthesis and measurements of materials have made clear that the two interactions are not mutually exclusive, and their interplay lies at the heart of many quantum materials, from transition-metal oxides\cite{Abramovitch2024_SVO,mirjolet2021_SVO,Coulter2026,Verdi2017,Reticcioli2019,Abramovitch2025_ephdmft}, molecular solid superconductors\cite{Gunnarson1997,Han2000,Capone2002,Capone2009,Faber2011,Nomura2015} and unconventional superconductors\cite{Yan2023_pnas,Braicovich2020,Gunnarsson2008}, to name a few.

This situation calls for a theoretical understanding of this interplay aiming at the development of a framework in which the two interactions can be treated on the same footing in the realistic description of materials\cite{Capone2010}.

In this light, minimal models that capture this competition, such as the Hubbard-Holstein model, remain central to theoretical investigations. A key simplification introduced by this model is the local nature of all the interactions. Despite that, it is an extremely rich playground whose computational investigation is far from trivial\cite{Grilli1994,Capone2010,Karakuzu2022,Han2020,Ohgoe2017,Johnston2013,Costa2020}. One of the reasons, that we highlight in the following is the presence of several independent energy scales and parameters that lead to different physical regimes. 

In the last decades, a central role 
has been played by Dynamical Mean‑Field Theory (DMFT)\cite{Georges1996}, which yields quantitatively reliable results for both electronic and phononic static and dynamical (spectral) properties \cite{Freericks1995,Koller_2004,Jeon2004,Capone2004,Werner2007,Sangiovanni2005,Sangiovanni2006,Paci2006,Bauer_2010,Sangiovanni2006afm}
 and does not require any perturbative assumption.

The Hubbard-Holstein model is characterized by a strong competition between the two interaction terms. Yet, non-trivial effects have been discovered. For example, DMFT studies of the Hubbard-Holstein model have focused on how electronic correlations strongly renormalize the effective electron-phonon coupling\cite{Abramovitch2025_ephdmft,Moghadas2025}, and how phonon effects manifest prominently at high frequencies, while leaving Fermi-liquid renormalizations of low-energy electronic behavior only weakly altered\cite{Sangiovanni2005,Coulter2025,Sangiovanni2006}.

Nonetheless, DMFT implementations for electron-phonon models can become computationally unfeasible, especially at low temperatures, for small phonon frequencies where a large number of phonons is excited, or in multi-orbital contexts.

As an alternative, slave-boson techniques~\cite{Coleman1984,Kotliar1986} — in particular the Rotationally Invariant Slave Boson (RISB) method~\cite{Lechermann2007} and the sibling Gutzwiller approximation\cite{Gutzwiller1965} offer a computationally efficient, variationally controlled, and symmetry-respecting framework for studying correlated electrons also in symmetry-broken phases~\cite{Isidori2009,Smit2025}. Despite the remarkable impact of slave-boson methods, conventional implementations are based on a mean-field decoupling that constrains the frequency structure of the self-energy, limiting their application especially in situations where an involved and rich frequency structure of the self-energy is expected as a result of competing interactions.

Recently, the advent of the ghost‑rotationally‑invariant slave boson (ghost-RISB) method~\cite{Lanata2022operatorial}, equivalent to the ghost-Gutzwiller approximation (ghost-GA)~\cite{Lanata2017bloch} at mean-field level, has dramatically improved and extended the capabilities of slave-boson approaches.
By introducing \textit{ghost} auxiliary orbitals, ghost-RISB extends the variational space of RISB, allowing it to mimic the dynamical self-energy structure typical of DMFT at a fraction of the computational cost.
Recent benchmarking~\cite{Lanata2017bloch,Frank2021anderson,Lee2023accuracy,Mejuto2023efficient} shows that ghost-RISB can reproduce quasiparticle weights, spectral functions, out-of-equilibrium dynamics~\cite{Guerci2023time}, and total energies of single-band and multiband Hubbard models, as well as in molecules~\cite{Mejuto2024molecules}, with increasing fidelity as the number of ghost orbitals grows.
This method has been successfully applied to solve a variety of strongly-correlated systems~\cite{Guerci2019mottexc,Giuli2025altermag,Tagliente2025spinons,pasqua2025qptop,Frank2021anderson,mejutozaera2026,kazemimoridani2026,Bellomia2026,pasqua2026fractionalized,tagliente2026direct} and it appears ideally suited for applying variational impurity solvers that may speed up quantum embedding calculations~\cite{Frank2024active,giuli2025linear}. Recently, it has been shown that in the limit of infinite ghosts this method becomes strictly equivalent to DMFT~\cite{giuli2026unification}.

So far, ghost-RISB (ghost-Gutzwiller) has only been applied to purely electronic Hamiltonians, leaving untouched the question of whether these upgraded slave-boson frameworks can be extended to systems coupled with bosonic degrees of freedom.
Such an extension is highly desirable in light of the ubiquity and importance of electron-phonon coupling in real materials.

Moreover, including phonons in DMFT faces some limitations such as not being able to access phononic quantities in Montecarlo solvers or being limited by the size of the bath in exact diagonalization solvers.

In this work we fill this gap by developing a generalization of ghost-RISB that incorporates dispersionless optical phonon modes locally coupled to arbitrary electronic modes. We benchmark this extension in the Hubbard-Holstein model, across a representative range of electron-electron and electron-phonon coupling strengths. We find excellent quantitative agreement with DMFT results present in the literature, capturing both the electron mass renormalization (quasiparticle weight suppression) and phonon-induced polaronic effects, as well as the competition between Mott and charge-localized or polaronic regimes and polaronic effects in superconductivity.

Importantly, our extended ghost-RISB achieves these results at a fraction of the computational cost of DMFT and with a smaller bath, similarly to what has been shown in the purely electronic case~\cite{Lee2023accuracy}, suggesting that it is a powerful and efficient tool for studying correlated materials where the electron-phonon interactions are significant and a truly dynamical treatment of the phonons becomes important.

We exploit this to characterize the superconducting state in the regime of strong electron-phonon coupling and small phonon frequency, proving that the zero-temperature order parameters drops for large coupling as a result of the formation of bipolarons.

Our work establishes extended ghost-RISB as a versatile and accurate framework for treating systems in which electronic and lattice interactions are intertwined, opening the door to systematic studies of complex regimes (e.g., strong correlation + strong e-ph coupling + realistic multi-orbital structure) that remain computationally challenging for fully dynamical methods.

The paper is structured as follows: in Section~\ref{sec:modelandmethod} we present the Hubbard-Holstein model and the formalism behind our generalized ghost-RISB.
In Section~\ref{sec:results}, we present the results of our calculations, which first serve as a benchmark against DMFT in the normal phase and then as a study of superconductivity in the bipolaronic regime. Thanks to our method, we are able to clearly identify the mechanism responsible of the suppression of superconductivity in the bipolaronic phase.
Finally, in Section~\ref{sec:conclusions} we draw the conclusions.  
\section{Model and Method} \label{sec:modelandmethod}

We study the physics of the fermionic polaron in the single-orbital Hubbard-Holstein model for which the Hamiltonian reads:

\begin{align} 
    \hat{H} =& \hat{H}_\text{el} + \hat{H}^\text{int}_{el}  + \hat{H}_\text{ph} + \hat{H}^\text{int}_\text{el-ph} \label{eq:full_hamiltonian} \\
    \hat{H}_\text{el} =& \sum_{\mathbf{R} \mathbf{R}^\prime,\sigma} t_{\mathbf{R} \mathbf{R}^\prime} c^\dagger_{\mathbf{R} \sigma} c_{\mathbf{R}^\prime \sigma} \nonumber \\
    \hat{H}^\text{int}_\text{el} =& \sum_{\mathbf{R}} \frac{U}{2} (\hat{N_{\mathbf{R}}}-1)^2 \nonumber \\
    \hat{H}_\text{ph} =& \sum_{\mathbf{R}} \omega_{0} b^\dagger_{\mathbf{R} \alpha} b_{\mathbf{R} } \nonumber \\
    \hat{H}^\text{int}_\text{el-ph} =& \sum_{\mathbf{R}} g ( b^\dagger_{\mathbf{R} } + b_{\mathbf{R}}) \hat{N}_\mathbf{R} \nonumber \\
\end{align}
where $c^\dagger_{\mathbf{R} \sigma}$ ($c_{\mathbf{R} \sigma}$) is the creation (annihilation) operator on site $\mathbf{R}$ for an electron with spin $\sigma$, $b^\dagger_\mathbf{R}$ ($b_\mathbf{R}$) is the creation (annihilation) operator of a phonon on site $\mathbf{R}$ and $\hat{N}_\mathbf{R}=\sum_{\sigma}c^\dagger_{\mathbf{R} \sigma} c_{\mathbf{R} \sigma}$ is the electrons' number operator at site $\mathbf{R}$.

The Hamiltonian is determined by the hopping matrix $t_{\mathbf{RR^\prime}}$, that we limit to nearest-neighbor sites,  the Hubbard repulsion $U$, the phononic frequency $\omega_0$, and the electron-phonon coupling $g$. The results will obviously also depend on the filling $n$. Two parameters can be defined to define different regimes of electron-phonon coupling, namely $\lambda = 2g^2/\omega_0$ and $\alpha = g/\omega_0$. In this work, we will mostly use $\lambda$. 

We choose the infinite-dimensional Bethe lattice, with a semi-circular non-interacting density of states, and we take its half-bandwidth $D$ as the unit of energy.
To solve this problem, we propose a generalization of ghost Rotationally-Invariant Slave-Bosons, an extension of Rotationally-Invariant Slave-Bosons that at the mean-field level corresponds to the recently developed ghost-Gutzwiller Approximation.
In the following, we derive the equations to treat local electron-phonon coupling adapted to the Hubbard-Holstein model, nonetheless, our generalized theory can take into account any number of local optical phonons coupled to local fermionic operators, as shown in Ref.~\cite{giuli2025phdthesis}.
We note that a previous generalization to include local phonons was theorized for the Gutzwiller Approximation in Ref.~\cite{Barone2006,Barone2008}, and we believe an equivalent formulation for the ghost-Gutzwiller Approximation is possible and would be equivalent to the one we present in this paper.
The nomenclature we use is largely inspired by Ref.~\cite{Lanata2022operatorial}.
We start mapping the local Fock space of the original problem, that we call \textit{original space}, to a vector space that is a subset of the Fock space of an enlarged system, that we called \textit{physical subspace}. The \textit{original space} is spanned by the following states on each site $\mathbf{R}$:
\begin{equation}
    | \Gamma_\mathbf{R} \rangle \otimes | {\Upsilon}_\mathbf{R} \rangle
\end{equation}
where
\begin{equation}
    | \Gamma_\mathbf{R} \rangle =[c^\dagger_{\mathbf{R} \uparrow}]^{m_\uparrow(\Gamma_{\mathbf{R}})} [c^\dagger_{\mathbf{R}\downarrow}]^{m_{\downarrow}(\Gamma_{\mathbf{R}})} |0\rangle_{el}
\end{equation}
are the electronic states with $\Gamma_{\mathbf{R}}\in \{ 0,...,2^{M}-1 \}$ and $m_i(\Gamma_{\mathbf{R}})$ is the $i$-th digit of the binary representation of $\Gamma_{\mathbf{R}}$. $M$ is the number of local fermionic modes, and $M=2$, accounting for spin up and down, for the single orbital problem.
The phononic states are instead:
\begin{equation}
    | {\Upsilon}_\mathbf{R} \rangle =  \frac{[b^\dagger_{\mathbf{R}}]^{\Upsilon}}{ \sqrt{\Upsilon !}} | 0 \rangle_{ph}
\end{equation}
and are generated by the operators $b^\dagger_{\mathbf{R}}$ applied to the phononic vacuum with $\Upsilon \in Z_{\geq0}$ non-negative integers counting the number of bosons in the state.
We will only consider translationally invariant problems, therefore we drop the subscript $\mathbf{R}$ since we consider the local Fock spaces to be all equivalent.

Now, as per any slave-particle approach, we need to define an \textit{enlarged space} and its \textit{physical subspace}, that is a subset of those states mapping to the \textit{original space}. For each site, consider a set of $\mathcal{P}=\mathcal{B} M  $ auxiliary Fermionic states generated by operators $f^\dagger_{a}$ with $a=1,...,\mathcal{P}$, where $\mathcal{B} \in \mathbb{Z}_+$ and a set of Bosonic modes generated by the operators $ \Phi_{\Gamma \Upsilon , m} $ with 
\begin{align}
\Gamma \in& \ \{ 0,...,2^{M}-1 \} \label{eq:RISB_imp_states}\\
m \in& \ S^\Gamma = \Big\{ m \in \{ 
0,...,2^{\mathcal{P}}-1 \} \ | N(m)-N(\Gamma) = \mathcal{M}  \Big\} \label{eq:RISB_bath_states}
\end{align}
and where $ N (\Gamma) $ and $ N (m) $ are respectively the number of electrons in states $\Gamma$ and $m$.
In general $\mathcal{M} $ can be any number, but from now on we will consider only $\mathcal{M} = \frac{M}{2} ( \mathcal{B}-1)$.
For the sake of simplicity, we will use the following notation $| \Gamma \rangle \otimes | {\Upsilon} \rangle  = | \Omega \rangle$ to include the fermionic and phononic states under a unique index $\Omega$.
We identify a set of states that compose the \textit{physical subspace} that will map to the Fock space of the original problem, defined as follows:

\begin{equation}
    | \widetilde{\Omega} \rangle = D^{-1/2}_{ \Omega} \sum_m \Phi^\dagger_{ \Omega ,m} [ f^\dagger_{1}]^{q_1(m)} ... [f^\dagger_{\mathcal{P}}]^{q_{\mathcal{P}}(m)} |0 \rangle_f | 0 \rangle_\Phi
\end{equation}

where:
\begin{equation}
    D_{ \Omega}= \sum_{m \in S^\Gamma} 1= \frac{\mathcal{P}!}{ (N(\Gamma)+\mathcal{M})! (\mathcal{P}-N(\Gamma)-\mathcal{M})!} \label{eq:ch2_Dgamma}
\end{equation}

Imposing the orthogonality and normalization of these states.

It can be proved~\cite{Lanata2022operatorial,giuli2025phdthesis} that these states are the only ones that satisfy a generalization of the so-called \textit{Gutzwiller constraints}:
\begin{align}
    &\sum_{\Omega,m} \Phi^\dagger_{\Omega m} \Phi_{ \Omega m} -1 =0\label{eq:GutCon_1} \\
    &\sum_{\Omega,mpq} \Phi^\dagger_{\Omega q} \Phi_{ \Omega m} [F^\dagger_{a}]_{m p} [F_{b}]_{pq}-f^\dagger_{a} f_{b} =0 \ \ , \ \ \forall \  a,b\label{eq:GutCon_2}
\end{align}

where $ [F^{(\dagger)}_{a}]_{mp} = \langle m | f^\dagger_{a} | p \rangle$ and $a,b \in \{1,...,\mathcal{B}M \}$.

From here on, the derivation is equivalent to that of Ref.~\cite{Lanata2022operatorial}, the only difference being that the impurity states $|\Omega\rangle$ contains also the phononic degrees of freedom.

We define a mapping between the \textit{original operators} acting on the \textit{original space} and the \textit{physical operators} acting on the \textit{physical subspace}. We start from the original creation operator $c^\dagger_\alpha$. The equivalent operator on the enlarged space will be given by:
\begin{equation}
    \widetilde{c}^\dagger_\alpha = \sum_a f^\dagger_a R_{a \alpha}
\end{equation}
to complete the mapping, the operators must satisfy the following condition:

\begin{equation} 
    \langle \widetilde{\Omega} | \widetilde{c}^\dagger_{\alpha } | \widetilde{\Omega}^\prime \rangle
    \overset{!}{=} \langle \Omega |c^\dagger_{\alpha } | \Omega^\prime \rangle 
    =[C^\dagger_{\alpha}]_{\Gamma \Gamma^\prime} \delta_{\Upsilon \Upsilon^\prime} \label{eq:ch2_tildec_cond}
\end{equation}
where we introduced the matrices $[C^\dagger_{\alpha}]_{\Gamma \Gamma^\prime}=\langle \Gamma | c^\dagger_\alpha | \Gamma^\prime \rangle $ in the last equality.

To satisfy the previous condition, we may use the following definition for the $\hat{R}$ operators:
\begin{equation}
    \hat{R}_{a \alpha} = \sum_{\Omega \Omega^\prime , m m^\prime} (\mathcal{N}_{\Omega \Omega^\prime , m m^\prime})^{-1/2} [C^\dagger_{\alpha}]_{\Omega \Omega^\prime } [F^\dagger_{a}]_{m m^\prime} \Phi^\dagger_{ \Omega m} \Phi_{ \Omega^\prime m^\prime} \label{eq:R}
\end{equation}

Where we define
\begin{align}
    \mathcal{N}_{\Omega \Omega^\prime,m m^\prime} =& [N(\Gamma)+\mathcal{M}][M+\mathcal{M}- N(\Gamma^\prime)] \nonumber \\
    =& [N(\Gamma)+\mathcal{M}][\mathcal{B}M-\mathcal{M}- N(\Gamma^\prime)]  \label{eq:ch2_Rfactor}
\end{align}

Since we know from equation \eqref{eq:RISB_bath_states} that $N(\Gamma) = \mathcal{M}-N(m)$, we can also write the normalization factor as:

\begin{equation}
\mathcal{N}_{\Gamma \Gamma^\prime, m m^\prime} = N(m)[\mathcal{B} \mathcal{M}-N(m^\prime) ]
\end{equation}

Following Ref. \citep{Lanata2022operatorial} one can think of generalizing this equation as follows:

\begin{widetext}
    
\begin{align}
    \hat{R}_{a \alpha} = \sum_{\Omega \Omega^\prime} \sum_{m m^\prime} \sum_{b} &(\mathcal{N}_{\Omega \Omega^\prime})^{-1/2} [C^\dagger_{\alpha} ]_{\Omega \Omega^\prime}[F^\dagger_{b}]_{mm^\prime} : \Phi_{\Omega m}^\dagger \Big[1+(\mathcal{N}_{\Omega \Omega^\prime}^{1/2}-1)]\sum_{\Omega^{\prime \prime}m^{\prime \prime}} \Phi^\dagger_{\Omega^{\prime \prime}} \Phi_{\Omega^{\prime \prime}} \nonumber \\
    &[(\hat{\mathbf{1}}-\hat{\Delta}_{p})^{-1/2}
    (\hat{\mathbf{1}}-\hat{\Delta}_{h})^{-1/2}
    ]_{ba} \Phi_{\Omega^\prime m^\prime} : \label{eq:R_2}
\end{align}
\end{widetext}

with
\begin{align}
    [\hat{\Delta}_{p}]_{ab} = \sum_{\Omega,pm} [F^\dagger_{a} F_{b}]_{mp} \Phi_{\Omega p}^\dagger \Phi_{ \Omega m} \label{eq:deltap_def} \\
    [\hat{\Delta}_{h}]_{ab} = \sum_{\Omega,pm} [F_{b} F^\dagger_{a} ]_{mp} \Phi_{\Omega p}^\dagger \Phi_{ \Omega m}\label{eq:deltah_def}
\end{align}

and clearly $[\hat{\Delta}_p]_{ab}=\delta_{ab}-[\hat{\Delta}_h]_{ab}$.
This is in principle sufficient to describe any fermionic operator since in the original theory these operators are functions of $c^\dagger_{\alpha}$ and $c_{\alpha}$.
However, for operators $\hat{O}_\mathbf{R}$  that are local in each cell $\mathbf{R}$, both fermionic and bosonic ones, an additional formulation can be given as:
\begin{equation} \label{eq:grisb_local_faithful}
    \hat{\widetilde{O}}_{\mathbf{R}} = \sum_{\Omega \Omega^\prime} [O_{\mathbf{R}}]_{\Omega \Omega^\prime} \sum_{m} \Phi^\dagger_{\mathbf{R}, \Omega m } \Phi_{\mathbf{R},\Omega^\prime m}
\end{equation}

where $ [O_{\mathbf{R}}]_{\Omega \Omega^\prime} = \langle \Omega_{\mathbf{R}} | \hat{O}_{\mathbf{R}} | \Omega^\prime_{\mathbf{R}} \rangle $. Thanks to property \eqref{eq:GutCon_1} it is easy to verify that:
\begin{equation}
       \langle \widetilde{\Omega}_{\mathbf{R}} | \hat{\widetilde{O}}_{\mathbf{R}} | \widetilde{\Omega}^\prime_{\mathbf{R}} \rangle = \langle \Omega_{\mathbf{R}} | \hat{O}_{\mathbf{R}} | \Omega^\prime_{\mathbf{R}} \rangle = [O_{\mathbf{R}}]_{\Omega \Omega^\prime} 
\end{equation}
this mapping also allows us to map the local terms of the Hamiltonians ($ \hat{H}_{\mathbf{R}}^\text{loc} $) to the following operator:
\begin{equation}
\hat{\widetilde{H}}^\text{loc}_{\mathbf{R}} = \sum_{\Omega \Omega^\prime} [ H^\text{loc}_{\mathbf{R}} ]_{\Omega \Omega^\prime} \sum_m \Phi_{\mathbf{R}, \Omega m}^\dagger \Phi_{\mathbf{R}, \Omega^\prime m} 
\end{equation}
Where $[ H^\text{loc}_{\mathbf{R}} ]_{\Omega \Omega^\prime}= \langle \Omega | H^\text{loc}_{\mathbf{R}} | { \Omega^\prime} \rangle$ and in the Hubbard-Holstein model:
\begin{align}
    H^\text{loc}_{\mathbf{R}} = & \sum_{\sigma}t_{\mathbf{R} \mathbf{R}}c^\dagger_{\mathbf{R}\sigma}c_{\mathbf{R}\sigma}+\frac{U}{2}(\hat{N}_{\mathbf{R}}-1)^2 \nonumber \\
    &+\omega_0 b^\dagger_\mathbf{R}b_{\mathbf{R}} +g(b^\dagger_\mathbf{R} +b_{\mathbf{R}})\hat{N}_\mathbf{R}
\end{align}
Therefore, completely encoding the phonons inside the additional bosonic fields.

With this last step, we built an enlarged theory that in the \textit{physical subspace} maps exactly to the original one.
In particular, the Hamiltonian of such a theory is given by:
\begin{equation}
    \hat{\widetilde{H}}_{\mathcal{B}} = \sum_{\mathbf{R}\neq\mathbf{R}^\prime} \sum_{\alpha \beta} t^{\alpha \beta}_{\mathbf{R} \mathbf{R}^\prime} \widetilde{c}^\dagger_{\mathbf{R} \alpha} \widetilde{c}_{\mathbf{R}^\prime \alpha} + \sum_{\mathbf{R}} \hat{\widetilde{H}}_{\mathbf{R}}^\text{loc} \label{eq:effective_Hamiltonian}
\end{equation}
where we would have to enforce the Gutzwiller constraints \eqref{eq:GutCon_1} and \eqref{eq:GutCon_2}.

With this formulation, the field theory is exact but still untractable. The standard approximation relies on two additional steps.
First, using a trial wavefunction made of a Slater determinant $| \psi_0 \rangle$ for the auxiliary fermions $f$ and a coherent wavefunction $| \phi \rangle$ for the slave-bosons $\Phi$, equivalent to taking the mean-field approximation on the bosonic fields, such that the total wavefunction is:
\begin{equation}
    | \Psi \rangle = | \psi_0 \rangle \otimes | \phi \rangle
\end{equation}
and second, promoting the matrices $\Delta_p=\mathbb{1}-\Delta_h$ and $R$ to be independent variables with the addition of two matrix Lagrange multipliers $\Lambda$ and $D$.
This would lead to a Lagrangian formulation that has exactly the same form as traditional ghost-RISB, but incorporates the phonons inside the slave-boson fields and therefore inside the embedding Hamiltonian.

In Appendix~\ref{app:lagrangian} we propose instead a derivation based on a purely field theory approach, which is completely equivalent to the traditional wavefunction-based approach but we believe would also be better suited for finite temperature generalization going in the direction of Ref.~\cite{giuli2026unification} and~\cite{Lanata2015finiteT}. We first promote $\hat{R}$ and $\hat{\Delta}_p$ to be independent variables, after this promotion the Lagrangian reads:

\begin{align}
      \mathcal{L}^\text{gRISB} =   \mathcal{L}^{\text{gRISB-}f}  &+ \sum_\mathbf{R}   \mathcal{L}^{ \text{gRISB-}\Phi}_{\mathbf{R}} \nonumber \\
      &+ \sum_\mathbf{R}   \mathcal{L}^\text{gRISB-mix}_{\mathbf{R}}
\end{align}
made of a fermionic part $\mathcal{L}^{\text{gRISB}-f}$, a slave-boson part $\mathcal{L}^{\text{gRISB}-\Phi}$ and a mixing part $\mathcal{L}^\text{gRISB-mix}$, the details are given in Appendix~\ref{app:lagrangian}.
The Lagrangian depends on a set of matrices $\Lambda_{\mathbf{R}}, R_{\mathbf{R}},\Lambda^c_{\mathbf{R}},D_{\mathbf{R}},\Delta_{\mathbf{R}},\Phi_{\mathbf{R}}$ on each site $\mathbf{R}$. From now on we consider translationally invariant solutions and we drop the subscript $\mathbf{R}$ on those matrices.
It has been shown~\cite{lanata2015,Lanata2022operatorial} that the mean-field decoupled theory for the bosons can be mapped into an interacting impurity problem for each correlated site $\mathbf{R}$, allowing for a drastic simplification of the self-consistency cycle coupling a non-interacting problem for the fermionic fields $\tilde{f}^{(\dagger)}_{\mathbf{R}\alpha}$ and the impurity problems.

The embedding Hamiltonian reads:
\begin{align}
\hat{H}_\text{emb} =& \hat{H}^\text{loc}[ c^\dagger,c,b^\dagger,b] + \sum_{ab} [\Lambda^c]_{ab} \tilde{f}_b \tilde{f}^\dagger_a \nonumber \\
&+ \sum_{a \alpha} [D_{a \alpha} c^\dagger_\alpha \tilde{f}_c   + H.c.]   \label{eq:H_emb}
\end{align}
Where $\hat{H}^\text{loc} $ is the local interacting Hamiltonian containing both the electron-electron and electron-phonon interaction as well as the local quadratic terms for electrons and phonons, 
$[\Lambda_{c}]_{ab}$ is the one-body Hamiltonian of the bath fermionic fields $\tilde{f}^{(\dagger)}_a$, and the hybridization between the impurity and the bath is:
\begin{equation}
    D_{ a \alpha} =  \sum_\mathbf{k}\sum_{\beta b} \epsilon_\mathbf{k}^{\alpha \beta} \Delta_{\mathbf{k},cb}  {R}^\dagger_{b \beta} \big[ \sqrt{\Big( \Delta({\mathbb{1}}-{\Delta})
     } \Big)^{-1}\big]_{ac} 
     \label{eq:V_eq}
\end{equation}
As shown in Ref~\cite{lanata2015} and in Appendix~\ref{app:lagrangian}, the saddle point equation with respect to the bosonic fields return an equation that can be interpreted as an eigenvalue equation of the embedding Hamiltonian, where the eigenstate is given by:

\begin{equation}
    |\Psi_{emb}\rangle = \sum_{\Omega n}   \Phi_{\Omega m}  s(m) | \Omega_\mathbf{R} \rangle \otimes [\hat{U}_{PH} |m_\mathbf{R} \rangle ] 
\end{equation}
where $s(m) = e^{i(\pi/2)N(m)[N(m)-1]}$ and $U_{PH}$ is a particle-hole transformation acting on the Hilbert space spanned by bath states $|n\rangle $.
Given this interpretation, the saddle point with respect to the first Gutzwiller constraint~\eqref{eq:GutCon_1} can be seen as the normalization condition of the state $|\Psi_{emb}\rangle$.

Moreover, eq.~\eqref{eq:R_2} can be rewritten as:
\begin{equation}
    \langle \hat{R}_{a \alpha}  \rangle = {R}_{a \alpha} = \sum_b \langle c^\dagger_\alpha \tilde{f}_b \rangle_{emb} \big[ \sqrt{\Big( \Delta({\mathbb{1}}-{\Delta})
     } \Big)^{-1}\big]_{ba}
     \label{eq:R_final}
\end{equation}
and the saddle point with respect to $\Delta$ gives:
\begin{align}
    [\Lambda^c+\Lambda]_{mn}= \Big[ &
 \sum_{\alpha c a} {R}_{\alpha c} \frac{\delta [\Delta(1-\Delta)]^{-1/2}_{ca} }{\delta 
 [\Delta]_{mn} }
 V_{a \alpha} + H.c.  \Big]
\label{eq:Delta_saddle_point}
\end{align}

The self-consistency cycle then consists in optimizing the matrices $\mathcal{R},\Lambda,V,\Lambda^c,\Phi,\Delta$ to realize the saddle point equations which are listed in Appendix~\ref{app:lagrangian}.

The self-consistency cycle we implement goes as follows:
\begin{enumerate}
    \item We start from a guess for matrices $\Lambda$ and ${R}$.
    \item We solve the auxiliary fermionic problem and compute the new $V$ from Eq.~\eqref{eq:V_eq} and the new $\Delta_{ab}=\langle f^\dagger_a f_b \rangle$. \label{itm:first_step_loop}
    \item We compute the new $\Lambda^c$ from Eq.~\eqref{eq:Delta_saddle_point}.
    \item We solve the impurity problem associated to the embedding Hamiltonian at eq.~\ref{eq:H_emb}, that is realizing the saddle point of the $\Phi$ fields.
    \item We update $\Delta_{ab}=\langle \tilde{f}_a \tilde{f}^\dagger_b \rangle$.
    \item We compute ${R}$ and $\Lambda$ respectively from Eq.~\eqref{eq:R_final} and \eqref{eq:Delta_saddle_point}.
    \item We check the convergence of $\Lambda$ and ${R}$ with the old parameter; if not converged, we go back to the point \ref{itm:first_step_loop}.
\end{enumerate}

We solve the embedding Hamiltonian at eq.~\eqref{eq:H_emb} with exact diagonalization. We typically truncate the phononic Hilbert space to 40 states.

\section{Results} \label{sec:results}

In this section, we will present some results to assess the accuracy and computational feasibility of ghost-RISB for the Hubbard-Holstein problem on the Bethe lattice, where a large number of DMFT studies are available. 
We will indeed first benchmark our method against DMFT results available in the literature,  and we will then apply this method to the study of superconductivity in the Hubbard-Holstein model, where we will exploit the accuracy of our approach to discuss the evolution of superconducting properties in terms of the phonon-displacement distribution function.

\subsection{DMFT benchmark}  \label{subsec:DMFT_benchmarks}

As we discussed above, the Hubbard-Holstein model depends on several control parameters, so that in this section we present some calculations that demonstrate that the ghost-RISB approach correctly reproduces: (i) the competition between electron-electron and electron-phonon coupling in determining the quasi-particle weight; (ii) the dependence on the filling $n$ and (iii) on the phonon frequency. We then address the phonon properties.

As far as the first two points are concerned, we compare with  DMFT results~\cite{Sangiovanni2005,Sangiovanni2006} using an exact diagonalization solver which, analogously to ghost-RISB, requires the solution of a discretized bath.

We emphasize that the strength of ghost-RISB is, by construction, its ability to accurately compute static observables. Spectral quantities can nevertheless be extracted from the renormalized Green's function, although they typically converge more slowly as a function of the number of ghost orbitals $\mathcal{B}$ with respect to static observables.

Therefore, the quasiparticle weight provides a meaningful benchmark for the validity of the approximation, being directly related to the dynamical structure of the self-energy at small frequencies according to:
\begin{equation}
    Z=\Big(1-\frac{\partial \text{Re}\Sigma(\omega)}{\partial \omega} \Big|_{\omega=0} \Big)^{-1}.
\end{equation}
Within the ghost-RISB, we have an analytical expression for the self-energy~\cite{giuli2026unification}: 
\begin{equation}
    \Sigma(\omega) = \omega 
-\Big[\mathcal{R}^\dagger \big(\omega \mathbf{1}-\Lambda \big)^{-1}\mathcal{R}\Big]^{-1}
+\mu
\end{equation}

We consider two cases, the half-filled case and the doped case, and compare how the effective mass is affected by the electron-phonon coupling.

\begin{figure}
    \centering
    \includegraphics[width=1.0\linewidth]{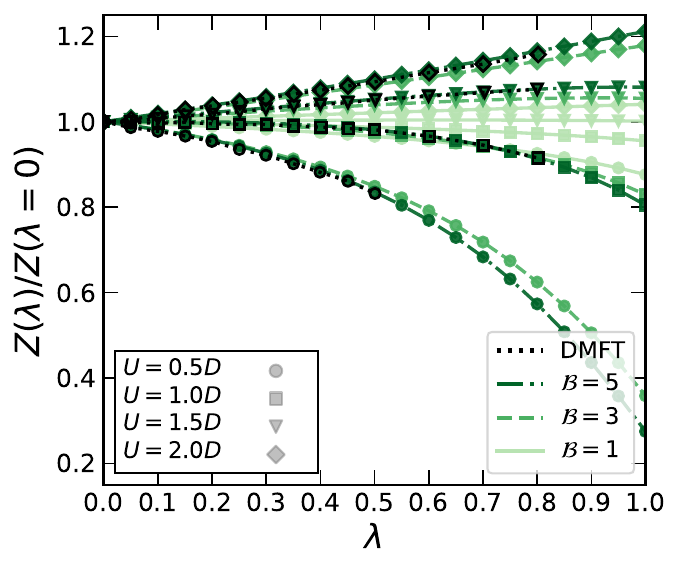}
    \caption{Effect of electron-phonon interaction on the quasi-
particle weight Z measured by the ratio $Z(\lambda)/Z(\lambda=0)$ for $\omega_0=0.2$ and $U=0.5D$ (circles), $U=1.0D$ (squares), $U=1.5D$ (triangles) and $U=2.0D$ (diamond).
The figure compares DMFT (black, dotted line) extracted from Ref.~\cite{Sangiovanni2005} with ghost-RISB using $\mathcal{B}=5$ (dark green, dash-dot line), $\mathcal{B}=3$ (green, dashed line) and $\mathcal{B}=1$ (light green, continuous line). }
    \label{fig:benchmark_half-filled}
\end{figure}

In Figure \ref{fig:benchmark_half-filled} we show the evolution, as a function of $\lambda=2g^2/\omega_0$, of the ratio between the phonon-corrected weight $Z(\lambda)$ and the value for $\lambda =0$. The calculations are performed for $\omega_0/D =0.2$ and for different values of the Hubbard interaction 
$U=0.5D$ (circles), $U=1.0D$ (squares), $U=1.5D$ (triangles) and $U=2.0D$ (diamond). We compare DMFT calculations with ghost-RISB with 
$\mathcal{B}=1$, 3 and 5. 
The DMFT data show a rather rich picture, which has been discussed in Ref. ~\cite{Sangiovanni2005}. For small $U$, the effect of the phonons is to reduce $Z$, which means increasing the correlations. This is essentially the conventional process of phonon-driven effective mass that evolves from a perturbative regime to a polaronic one\cite{}. When $U$ is increased to $U=D$, we find a strong competition between the two interactions which result is a near balance testified by a rather flat $Z(\lambda)$ before a polaronic regime is reached when $\lambda$ exceeds $U$. On the other hand, for large $U$, the effect of $\lambda$ is to reduce the degree of correlation (increasing $Z$). This is easily interpreted in terms of a reduced effective repulsion which equals $U_{eff} = U -\lambda$ in the antiadiabatic limit $\omega_0/D \gg 1$.

It is clear from the figure that the $\mathcal{B}=5$ values (dark green) are already indistinguishable from the DMFT data (black), which were obtained for a larger bath.  The comparison shows that the $\mathcal{B}=3$ values (green) are in remarkable qualitative and quantitative agreement with DMFT, despite deviating slightly. This testifies that the ghost-RISB is able to accurately capture the competition between electron-electron and electron-phonon interactions.

On the other hand, the $\mathcal{B}=1$ values, which correspond to the standard RISB method, show important deviations from the DMFT ones, and it does not capture the evolution between different regimes of interplay. 
This confirms that the ghost orbitals are fundamental to describe a retarded interaction, like the attraction mediated by phonons, and the $\mathcal{B}=1$ case presents only some qualitative effect as it was shown in previous approaches based on a generalized Gutzwiller wavefunction~\cite{Barone2006,Barone2008}.

\begin{figure}[h!]
    \centering    \includegraphics[width=\linewidth]{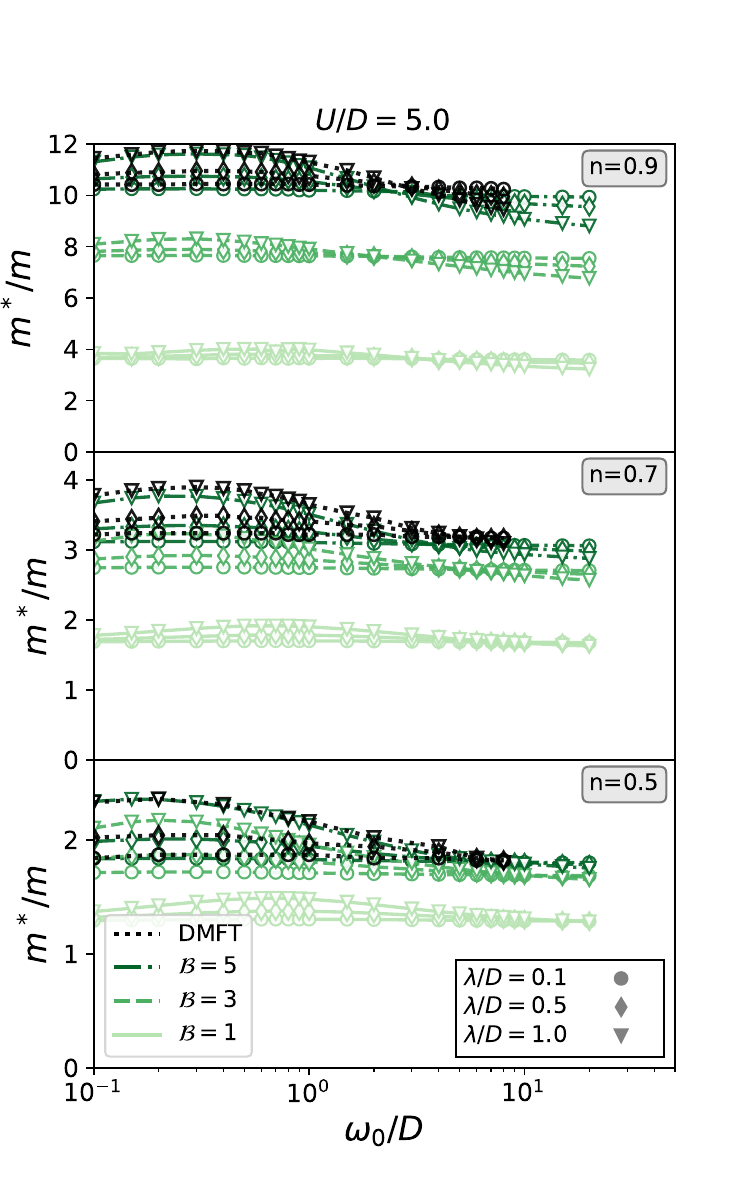}
    \caption{Mass renormalisation as a function of the phonon frequency for $n=0.9$ (top panel), $n=0.7$ (central panel) and $n=0.5$ (bottom panel) at fixed $U=5D$.  In each panel, we plot $\mathcal{B}=1$ (light green, continuous line), $\mathcal{B}=3$ (green, dashed line) and $\mathcal{B}=5$ (dark green, dash-dot line), together with DMFT (dotted gray lines). The markers indicate $\lambda=1.0D$ (triangle), $\lambda=0.5D$ (diamond) and $\lambda=0.1D$ (circle). The agreement increases with doping and with the number of ghosts. $\mathcal{B}=5$ is in perfect agreement with DMFT data from Ref.\cite{Sangiovanni2006}. }
    \label{fig:benchmark_doped}
\end{figure}

In Figure \ref{fig:benchmark_doped} we focus on the dependence on filling. At the same time, we address the dependence on the phonon frequency plotting the results as a function of $\omega_0/D$. The choice to highlight the role of filling together with the phonon frequency is suggested by the analysis of Ref. \cite{Sangiovanni2006}, which shows that the degree of adiabaticity has much stronger signatures for doped systems than at half-filling.

We fix $U=5D$, and consider three values of filling:  $n=0.9$ (top panel), $n=0.7$ (central panel) and $n=0.5$ (bottom panel).
We consider, the effective electron-phonon interaction $\lambda=0.1$  (circles), $\lambda=0.5$ (diamonds), $\lambda=1.0$ (triangles) and we study how the mass renormalization, that is the inverse of the quasiparticle weight $m^*/m=1/Z$, is affected by the phononic frequency for different numbers of ghost orbitals $\mathcal{B}=5$ (dark green, dash-dot line) $,\mathcal{B}=3$ (green, dashed line) and $\mathcal{B}=1$ (light green, continuous line), compared to DMFT (black, dotted line).
Similarly to the half-filled case we notice that ghost-RISB is already almost indistinguishable from DMFT results for $\mathcal{B}=5$ and, as previously, $\mathcal{B}=3$ captures the qualitative behavior and is only in partial quantitative agreement, while $\mathcal{B}=1$ has a smaller enhancement effect.
Moreover, we also notice that the agreement with DMFT increases with increasing doping.

Lastly, we address the extent to which the method can capture the properties of the phonons. As we mentioned in passing, the most direct signature of electron-phonon coupling is the formation of polarons and bipolarons, which have been widely studied within DMFT\cite{Ciuchi1997,Benedetti1998,Capone2003,Capone2006_pol}. This crossover can be effectively pinpointed by the probability distribution function (PDF) of the phonon field\cite{Capone2006_pol}. Within the ghost-RISB formalism, it can be expressed as
\begin{align}
    P(x) =&  \sum_{m,\,\Gamma} \langle m,\Gamma ,x|\Psi \rangle  \langle \Psi| m, \Gamma , x\rangle \nonumber \\
=&
\sum_{m,\,\Gamma}
\left|
\sum_\Upsilon \phi_n(x)\psi_{m,\Gamma,\Upsilon}
\right|^2
\end{align}
Where $| \Psi \rangle = \sum_{m,\Gamma,\Upsilon} \psi_{m,\Gamma,\Upsilon} |m , \Gamma,\Upsilon\rangle$ is the ground state of the embedding Hamiltonian and $|m\rangle,|\Gamma \rangle ,|\Upsilon \rangle$ are respectively bath, impurity and phononic states.
$\phi_\Upsilon (x)$ is the wavefunction of the $\Upsilon$-th quantum harmonic oscillator bound state for the displacement operator $x=\frac{b^\dagger + b}{\sqrt{2}}$ on a given site \cite{millis1996, ciuchi1999, Capone2006_pol}. This function exhibits a single peak at $x=0$ for small $\lambda$, but may develop a bimodal structure when the el-ph coupling exceeds the crossover value to the polaronic regime. A bimodal PDF indicates a substantial polarization of the lattice and is associated with large probabilities of singly and doubly occupied sites as a consequence of the strong attractive interaction \cite{Capone2006_pol, Sangiovanni2006}.

\begin{figure}[h!]
    \centering
    \includegraphics[width=1.0\linewidth]{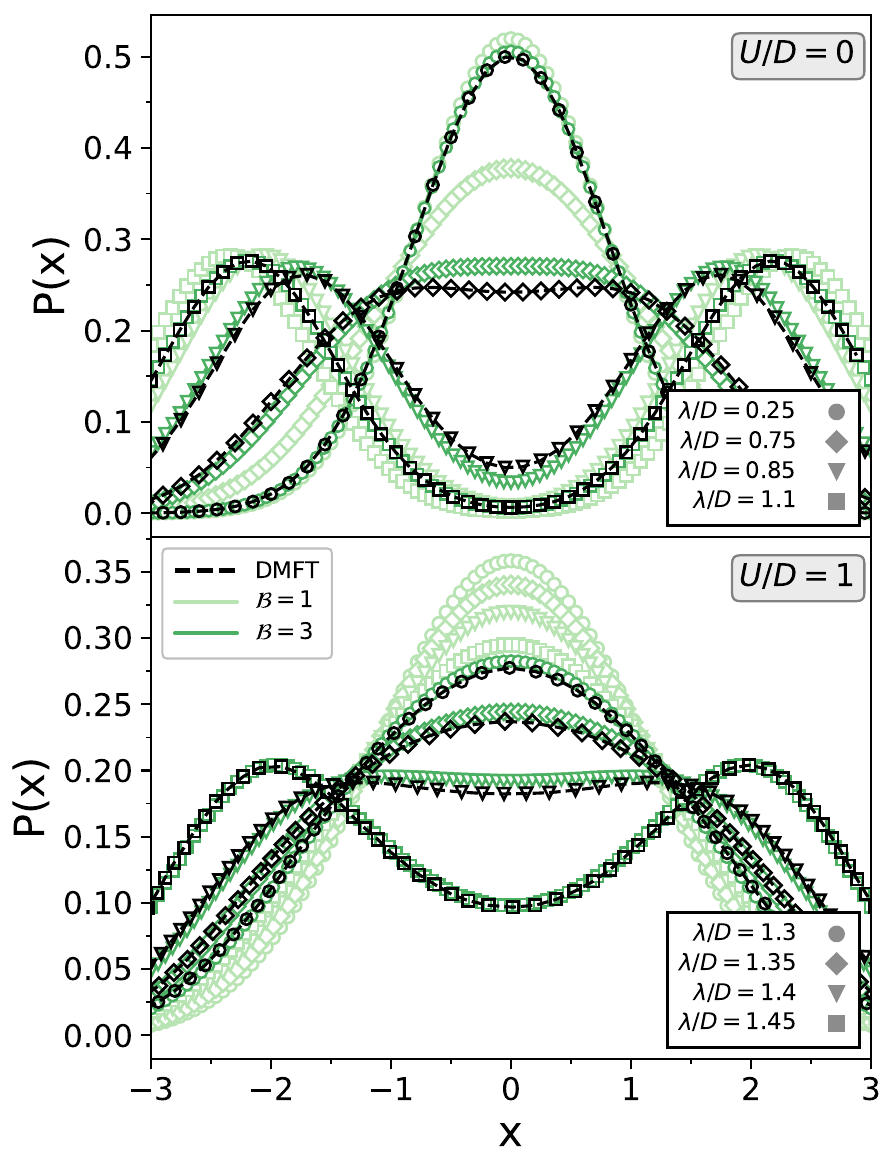}
    \caption{Phonon probability distribution functions for different values of $\lambda$. We compare the case of the half-filled Holstein model (upper panel) and the Hubbard-Holstein model at $U/D=1$ (lower panel), both with a characteristic frequency of $\omega_0/D=0.2$. The calculations are for $\mathcal{B}=1$ and $\mathcal{B}=3$ and are compared with DMFT data extracted from \cite{Capone2006_pol, SangiovaniDiss}.}
    \label{fig:PDF}
\end{figure}

In Figure ~\ref{fig:PDF} we show the generic evolution of the PDF with increasing $\lambda$ for the Hubbard-Holstein model, both at vanishing Hubbard interaction, $U=0D$ (upper panel), and at finite interaction, $U=1D$ (lower panel). We focus on the half-filled system at a fixed phonon frequency of $\omega_0=0.2 D$ and compare our ghost-RISB results for $\mathcal{B}=1$ (dotted lines) and $\mathcal{B}=3$ (solid lines) to DMFT data (bold solid lines) extracted from Ref.~\cite{Capone2006_pol,SangiovaniDiss}. In both models, the values of the el-ph interaction are chosen such that a sufficiently large $\lambda$ yields a bimodal PDF, indicating that the system is in the polaronic regime. Overall, we find remarkable quantitative agreement between the DMFT and $\mathcal{B}=3$ data. However, similarly to the examples discussed above, the $\mathcal{B}=1$ results show significant deviations, although, as can be seen for the two largest $\lambda$ points for $U=0D$, the method is able to capture polaron formation at least qualitatively. In contrast, for finite $U$ and for all $\lambda$ values considered here, the crossover to the polaronic regime is not observed by $\mathcal{B}=1$ and is only captured with $\mathcal{B}=3$ ghost orbitals. This highlights the limitation of the conventional RISB, while the ghost-RISB approach, even with only $\mathcal{B}=3$ sites, is sufficient to correctly capture how lattice properties are affected by the interplay between electrons and phonons.

\begin{figure}[h!]
    \centering
    \includegraphics[width=1.0\linewidth]{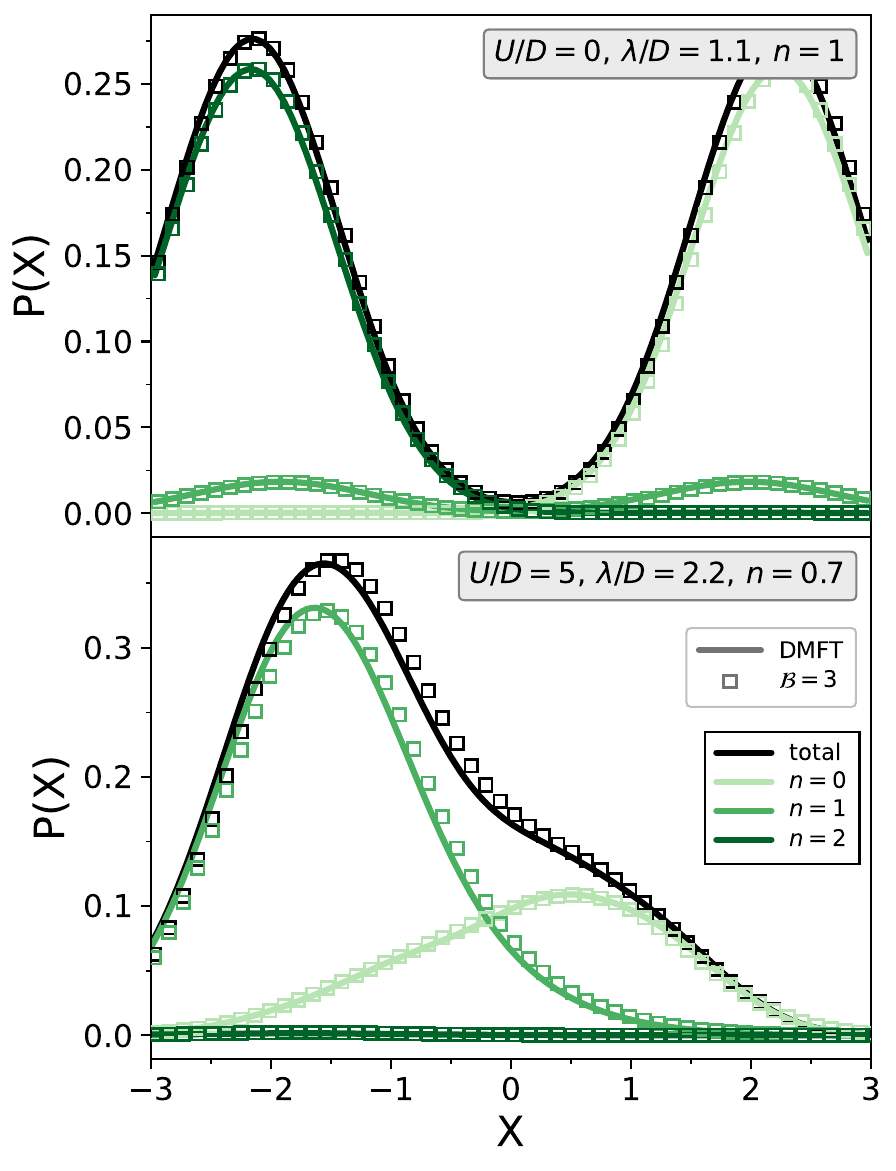}
    \caption{Total and charge-resolved phonon probability distribution functions projected to the empty, single, and double-occupied charge sectors. We compare the half-filled Holstein model (upper panel) at $\lambda/D=1.1$ and the doped Hubbard-Holstein model at $U/D=5, \, \lambda/D=2.2$ (lower panel) at a filling of $n=0.7$, both with a characteristic frequency of $\omega_0/D=0.2$. The calculations are done for $\mathcal{B}=3$ and compared with DMFT data at the same parameter values.}
    \label{fig:PDF_cr}
\end{figure}

In order to highlight the entanglement between phonon distortion and the electronic states, it can be useful to compute charge-resolved PDFs.
In our approach, such a conditional probability distribution can be computed for a given subset of states $\alpha$ belonging to local configurations $S$:
\begin{align}
    P_S(X) =&  \sum_{m} \sum_{\Gamma \in S} \langle m,\Gamma ,x|\Psi \rangle  \langle \Psi| m, \Gamma , x\rangle \nonumber \\
=&
\sum_m\sum_{\Gamma \in S}
\left|
\sum_\Upsilon \phi_n(x)\psi_{m,\Gamma,\Upsilon}
\right|^2
\end{align}
The charge sectors are $S=0,1,2$ which correspond respectively to the set of empty states ( $\{|0\rangle \}$, $S=0$ ) single occupied ( $\{ |\uparrow \rangle,| \downarrow \rangle \}$ , $S=1$ ) and doubly occupied ( $\{ |\uparrow \downarrow \rangle \}$ , $S=2$ ).

In Figure ~\ref{fig:PDF_cr} we demonstrate this charge-resolved PDF for two exemplary parameter sets and compare with the respective DMFT solutions, obtained with an exact-diagonalization solver based on the \textit{EDIpack} \cite{Crippa2025SPC, Crippa2025SPCa} software-package. For both cases, we fix the phonon frequency to $\omega_0/D=0.2$ and choose $\mathcal{B}=3$ for the ghost orbitals. The upper panel showcases the charge-resolved PDFs (shades of green) and the total distribution (black) for the polaronic regime of the half-filled Holstein model at an interaction value of $\lambda/D=1.1$. We note a remarkable agreement between the ghost-RISB and DMFT data for all three charge sectors and the total PDF. Evidently, the two main features of the full PDF are associated with the large number of empty and doubly occupied sites, while the $n=1$ charge sector plays a negligible role. In the lower panel instead, we demonstrate the PDFs corresponding to the doped Hubbard-Holstein model with $n=0.7$ filling at a fairly large Hubbard interaction of $U/D=5$. Also here, the results are in good qualitative agreement with the associated DMFT data. At the chosen el-ph interaction strength of $\lambda/D=2.2$, the emergence of a secondary peak in the PDF indicates that the system is at the verge of the polaronic regime. However, in contrast to the pure Holstein case or the examples of Figure ~\ref{fig:PDF}, the bimodality here originates from empty and singly occupied sites, as the large Hubbard interaction penalizes double occupancies. This clearly distinguishes the polaronic state from the bipolaronic one found at half-filling.

\subsection{Superconductivity}

In this section we extend our study to the s-wave superconducting phase of the model. In the absence of electron-electron interaction, the Holstein model shows an s-wave superconducting low-temperature state for any non-zero phonon frequency and electron-phonon coupling. Under very general circumstances, we expect the s-wave state to survive the inclusion of a finite $U$ as long as the effective interaction remains attractive. 

In this section, we show that the ghost-RISB is able to capture the superconducting phase and its non-trivial properties as a function of the interaction strength. In particular, we present some new insight on the strong-coupling regime, where the superconducting state is expected to present bipolaronic features, associated with large lattice distortions. 

From a technical point of view, we use a mapping that allows us to study superconductivity without breaking the particle-conservation constraint. In particular, we exploit a unitary particle-hole transformation which acts only on the down spins $c_{i\downarrow} \to c^\dagger_{i\downarrow}$. In this way the s-wave order amplitude $\Delta = \frac{1}{N}\sum_i \langle c^\dagger_{i\uparrow} c^\dagger_{i\downarrow}+c_{i\downarrow}c_{i\uparrow} \rangle$ (without loss of generality we assume it real) maps onto the ferromagnetic magnetization along the x direction $m_x = \frac{1}{N}\sum_i \langle c^\dagger_{i\uparrow} c_{i\downarrow}+c^{\dagger}_{i\downarrow}c_{i\uparrow} \rangle$.
The transformed local interaction reads:
\begin{align}
    U(n_\uparrow-\frac{1}{2})(n_\downarrow-\frac{1}{2})  &\rightarrow -U(n_\uparrow-\frac{1}{2})(n_\downarrow-\frac{1}{2}) \\
    g(b^\dagger+b)( n_\uparrow + n_\downarrow ) &\rightarrow g(b^\dagger+b)( n_\uparrow - n_\downarrow +1) 
\end{align}
the spin-down hopping changes sign under the transformation, so that the two spin species acquire opposite dispersions in the transformed representation. We performed calculations in the new basis, translating the results for the magnetic ordering of the new model into information about the superconducting phase of the Hubbard-Holstein model.

\begin{figure}[h!]
    \centering
    \includegraphics[width=\linewidth,trim={0 10 0 0}, clip]{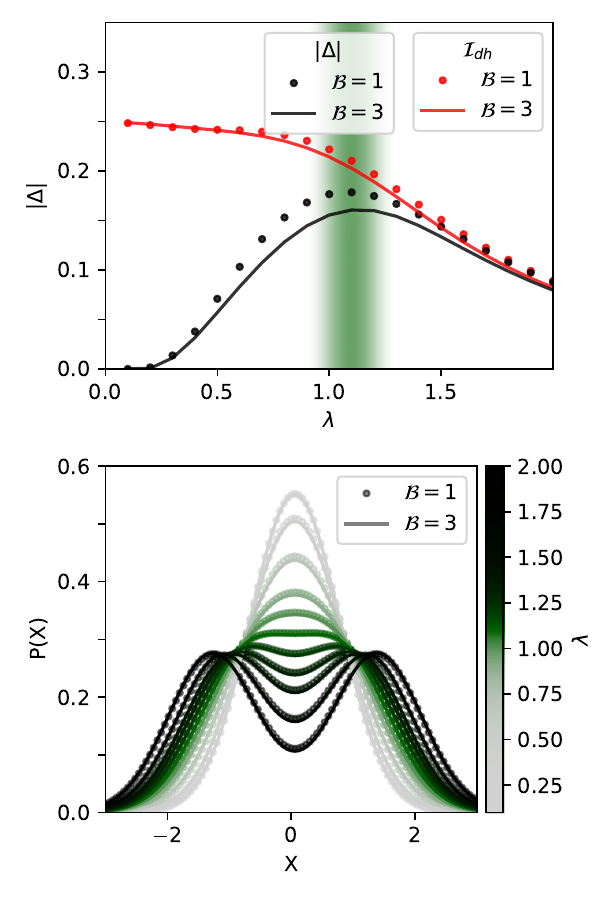}
    \caption{Holstein model ($U/D=0$ and $\omega_0/D=1.0$) plot of (a) superconducting order parameter (black) and double-holon overlap $\mathcal{I}_{dh}$ (red) for $\mathcal{B}=1$ (dots) and $\mathcal{B}=3$ (line) and as a function of the effective attraction $\lambda$. In green the area at which we identify the polaronic crossover.
    (b) Phononic distribution function showed for values of $\lambda/D$ between 0.0 (light gray) and 2.0 (black) passing through the polaronic crossover (green) around $\lambda=1.1$ using $\mathcal{B}=1$ (dots) and $\mathcal{B}=3$ (line).  
    }
    \label{fig:SC_Btest}
\end{figure}

In the limit of large phonon frequency, the Hubbard-Holstein model maps onto a Hubbard model with interaction strength $U_\text{eff}=U-\lambda$, which means an attractive Hubbard model if $\lambda > U$. The latter model is known to have a superconducting ground state for every value of $\vert U_\text{eff} \vert$. Within DMFT~\cite{Toschi2005_PRB,toschi2005_NJP} it has been shown that the amplitude of the order parameter is an increasing function of $\vert U_\text{eff}\vert$ which saturates to a constant in the infinite-$U_\text{eff}$ limit. However, in the large-$U_\text{eff}$ regime the critical temperature decays like $1/U_\text{eff}$. This is the Bose-Einstein Condensation (BEC) regime, in which superconductivity takes place as a condensation of preformed pairs of large amplitude. The evolution as a function of the attractive strength is smooth, and it describes the so-called BCS-BEC crossover.

When the phonon frequency is finite, and the phonons retain their intrinsic dynamics, the BEC regime becomes better described in terms of bipolarons, bound pairs of polarons created by the large electron-phonon coupling. It has been reported within DMFT\cite{Murakami2013}, that in the polaronic regime not only the critical temperature, but also the amplitude of the order parameter decay as the attraction becomes very large, signaling a distinctive difference between the BEC regime of the attractive Hubbard model and the bipolaronic regime of the Holstein (or of the Hubbard-Holstein) model. In this section we explore these aspects, identifying a physical picture justifying the reduction of the order parameter in the bipolaronic regime~\cite{Murakami2013,Nosarzewski2021,Zhang2023}. We notice in passing that different results have been reported for non-local electron-phonon coupling, where high-temperature bipolaronic superconductivity has been proposed~\cite{Sous2023,Zhang2023}.

First, we test how different values of $\mathcal{B}$ affect the description of the superconducting state. In Fig.~\ref{fig:SC_Btest} we study using $\mathcal{B}=1$ (dots) and $\mathcal{B}=3$ (lines) the superconductive order parameter Fig.~\ref{fig:SC_Btest}(a) and the phonon-displacement distribution function Fig.~\ref{fig:SC_Btest}(b)  for the Holstein model ($U=0$) for different values of $\lambda$ between $\lambda /D=0$ (light gray) and $\lambda /D=2$ (black).
Somewhat in contrast with the metallic solutions we have shown in the previous section, in the superconducting state, the $\mathcal{B}=1$ results (Standard RISB) are already very close to the $\mathcal{B}=3$, suggesting a faster convergence with bath size in symmetry-broken phases. This result can be connected with the popular notion that broken-symmetry solutions of correlated models display weaker correlation effects if compared with symmetric solutions~\cite{Bellomia2026}. For this reason, the simpler $\mathcal{B}=1$ calculations are sufficient to capture the correct results. However, we believe that further investigations of various symmetry breakings are in order to validate and/or disprove this statement.

Turning to the physical content of our results, we clearly observe a drop on $\Delta$ for large values of $\lambda$, analogously to Ref. \cite{Murakami2013}. In the following we exploit the ghost-RISB to identify the physical origin of this behavior. In particular we show that the reduction of $\Delta$ can be traced back to the development of the polaronic behavior (vertical green shade)  as measured by the onset of bimodality on the probability distribution function (PDF) of the phononic displacement operator shown in  Figure~\ref{fig:SC_Btest}(b) where the PDFs on the polaronic crossover are green.

Indeed, comparing the data of the two panels, the maximum of $\Delta$ almost coincides with the value of $\lambda$ for which $P(X)$ becomes bimodal (green vertical shaded line in panel (a) and green curves in panel (b)).

The connection can be established by proving that the superconducting order parameter is bounded from above by the overlap between the charge-resolved phononic PDFs, more precisely:

\begin{equation}
    |\Delta| \leq \int dx \sqrt{P_0(X)}  \sqrt{P_{2}(X)} = \mathcal{I}_{dh}
    \label{eq:orthogonality}
\end{equation}

where $P_0(X)$ and $P_{2}(X)$ are respectively the phononic probability distribution functions of the displacement operator $X=\frac{b^\dagger + b}{\sqrt{2}}$ on a given site when the electronic configuration on that site is, respectively, an empty or a doubly-occupied site, as shown in section~\ref{subsec:DMFT_benchmarks}. The derivation is straightforward, but we prefer to report it in Appendix B to make this section more readable. 

The right-hand side of Eq. (\ref{eq:orthogonality}) is the overlap between $P_0(X)$ and $P_{2}(X)$, i.e., the Bhattacharyya coefficient between the charge-resolved phonon distributions.
The bimodal distribution associated with the bipolaronic state is precisely due to the splitting between these two particle distributions. Hence, $I_{dh}$ is essentially independent on $\lambda$ when $P(X)$ is unimodal, and it starts to drop when we enter the bipolaronic regime.

In Figure~\ref{fig:SC_Btest}(a) we show how the suppression of the order parameter tracks the decrease of its upper bound $\mathcal{I}_{dh}$ associated with bipolaron formation.

\begin{figure}[h!]
    \centering
    \includegraphics[width=1.0\linewidth,trim={0 10 10 0}, clip]{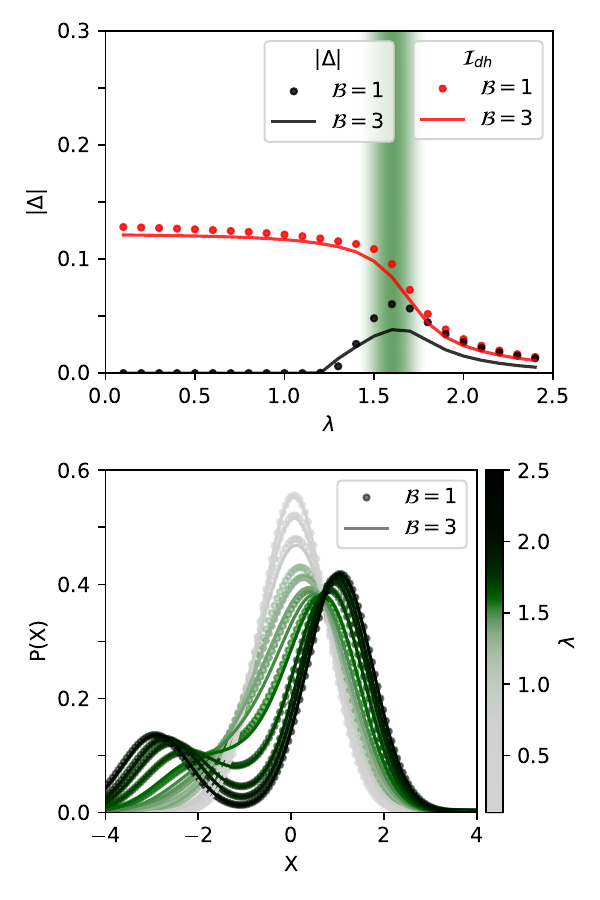}
    \caption{Hubbard-Holstein model ($U/D=1.0$ and $\omega_0/D=1.0$) plot of (a) superconductive order parameter (black) and double-holon overlap $\mathcal{I}_{dh}$ (red) for $\mathcal{B}=1$ (dots) and $\mathcal{B}=3$ (line) and as a function of the effective attraction $\lambda$. In green the area at which we identify the polaronic crossover.
    (b) Phononic distribution function shown for values of $\lambda/D$ between 0.0 (light gray) and 2.0 (black) passing through the polaronic crossover (green) around $\lambda=1.6$ using $\mathcal{B}=1$ (dots) and $\mathcal{B}=3$ (line).}
    \label{fig:bipol_SC_doped}
\end{figure}

This result is completely general, i.e., it also works in the presence of Coulomb repulsion. In Fig.\ref{fig:bipol_SC_doped} we show the same quantities as in Fig.\ref{fig:SC_Btest}, for the Hubbard-Holstein model at quarter-filling $n=0.5$ with $U/D=1.0$, $\omega_0/D=0.5$. In this doped case, we notice that the reduction of the overlap $\mathcal{I}_{dh}$ is much faster and dramatic, leading to a more pronounced drop of the superconducting order parameter.

This stronger effect can be understood by noticing that in a more adiabatic (low-frequency) bipolaronic regime, the two distributions become increasingly separated, giving rise to a kind of phonon-induced orthogonality between the corresponding local charge configurations. Electronic configurations that could be connected by an instantaneous interaction are no longer connected when the adiabatic character of the phononic motion is taken into account.

As a consequence, the superconducting order parameter is suppressed even when local pairing correlations remain sizable (i.e., the number of double-occupied sites is large). 

This mechanism can be interpreted as a kind of Franck--Condon effect in which the holon and doublon configurations play the role of two different electronic states. The reduction of coherent doublon-holon superposition induced by bipolaronic dressing of the two states causes the drop of the order parameter, which cannot be understood solely in terms of quasiparticle mass renormalization or static fermionic susceptibilities.
\section{Conclusions} \label{sec:conclusions}

In this work, we presented a formulation of the ghost-rotationally invariant slave-boson method designed to include local electron-phonon interactions.
By extending the ghost-orbital construction to systems coupled to local bosonic degrees of freedom, the method captures dynamical self-energy effects and phonon-induced renormalizations beyond conventional static slave-boson approaches while maintaining a computational cost significantly lower than dynamical mean-field theory.
Benchmark calculations for the Hubbard-Holstein model demonstrate that the method quantitatively reproduces the main dynamical mean-field theory features across a broad range of interaction strengths and phonon frequencies, including quasiparticle renormalization, polaronic crossover physics, and the competition between Mott and bipolaronic localization.
In particular, we showed that the inclusion of ghost orbitals is essential to correctly describe the low-frequency adiabatic regime, where dynamical phonon effects become dominant, and the conventional mean-field of the slave bosons (analogue to the Gutzwiller approximation) fails.
We then employed the method to investigate the superconducting phase of the Hubbard-Holstein model. Our results indicate that the suppression of the superconducting order parameter in the adiabatic regime is not associated with a reduction of the effective pairing interaction, but rather with a phonon-induced orthogonality between the empty and doubly occupied local charge sectors.
By expressing the anomalous expectation value in terms of the overlap between charge-resolved phonon distributions, we identified a Franck-Condon reduction of coherent doublon-holon mixing as the microscopic mechanism underlying the weakening of superconducting coherence in the bipolaronic regime.

The present formulation opens several perspectives for future work. The low computational cost of the method makes it particularly suitable for the rapid exploration of phase diagrams in multiorbital electron-phonon systems, nonequilibrium extensions, and materials-oriented applications where the simultaneous treatment of strong electronic correlations and dynamical lattice fluctuations remains computationally challenging.
Moreover, the present formulation naturally interfaces with already existing impurity solvers for electron-phonon systems, such as EDIpack~\cite{amaricci2022CPC,Crippa2025SPCa,Crippa2025SPC}, or software capable of treating both bosonic and fermion fields as Pomerol~\cite{Antipov2015} or Tensor Network softwares~\cite{ITensor2022,TensorKit2026,zhai2023block2}, enabling the direct integration of established packages within the ghost-RISB framework for electron-phonon systems.  

\section*{Acknowledgements} \label{sec:acknowledgements}

We are thankful to S. Ciuchi, A. Georges and G. Bellomia for insightful discussions.
We acknowledge financial support from the National Recovery and Resilience Plan PNRR MUR Project No. CN00000013-ICSC, MUR
Project No. PE0000023-NQSTI  and PRIN 2022 (Prot. 20228YCYY7). E.M. acknowledges support from the Austrian Science Fund (FWF) through the grant 10.55776/I5868 (Project P1 of the research unit QUAST, for5249, of the German Research Foundation, DFG)
The Flatiron Institute is a division of the Simons Foundation.

\bibliography{biblio}

\appendix
\begin{widetext}
\section{Lagrangian of the electron-phonon ghost-RISB and embedding formulation} \label{app:lagrangian}

The ghost-RISB method is based on the mapping of the interacting problem into the effective Hamiltonian at Eq.~\eqref{eq:effective_Hamiltonian} restricted within the physical subspace as given by the Gutzwiller constraints \eqref{eq:GutCon_1} and \eqref{eq:GutCon_2}.

We simplify the search of a solution by performing a mean-field decoupling of bosons and fermions and performing a local approximation on the bosons, therefore effectively decoupling at mean-field level also bosons belonging to different sites. 

The partition function is represented as the functional integral:

\begin{equation}
    Z=\int \mathcal D[\bar f,f],
\mathcal D[\bar\Phi,\Phi],
\mathcal D E,
\mathcal D\Lambda;
e^{-S}
\end{equation}

where $S=\int_{0}^{\beta}d\tau 
\mathcal L^{gRISB}(\tau)$ and:

\begin{align}
    \mathcal{L}^{\mathrm{gRISB}} (\tau) =& \sum_{\mathbf{R},a} \bar{f}_{\mathbf{R}a}(\tau) \partial_{\tau} f_{\mathbf{R}a}(\tau) \nonumber \\
    &+ \sum_{\mathbf{R},\Omega,n} \bar{\Phi}_{\mathbf{R},\Omega n}(\tau) \partial_{\tau} \Phi_{\mathbf{R},\Omega n}(\tau) \nonumber \\
    &- \sum_{\mathbf{R},\mathbf{R}'} \sum_{\alpha,\beta} t_{\mathbf{R}\mathbf{R}'}^{\alpha\beta} \left[ \sum_{a} \bar{f}_{\mathbf{R}a}(\tau) R_{\mathbf{R},a\alpha} \bigl[ \bar{\Phi}_{\mathbf{R}}(\tau), \Phi_{\mathbf{R}}(\tau) \bigr] \right] \nonumber \\
    &\qquad\qquad\times \left[ \sum_{b} R_{\mathbf{R}',b\beta}^{\dagger} \bigl[ \bar{\Phi}_{\mathbf{R}'}(\tau), \Phi_{\mathbf{R}'}(\tau) \bigr] f_{\mathbf{R}'b}(\tau) \right] \nonumber \\
    &+ \sum_{\mathbf{R}} \sum_{\Omega,\Omega'} \left[ H_{\mathbf{R}}^{\mathrm{loc}} \right]_{\Omega\Omega'} \sum_{n} \bar{\Phi}_{\mathbf{R},\Omega n}(\tau) \Phi_{\mathbf{R},\Omega' n}(\tau) \nonumber \\
    &+ \sum_{\mathbf{R}} E_{\mathbf{R}}(\tau) \left[ \sum_{\Omega,n} \bar{\Phi}_{\mathbf{R},\Omega n}(\tau) \Phi_{\mathbf{R},\Omega n}(\tau) - 1 \right] \nonumber \\
    &+ \sum_{\mathbf{R}} \sum_{a,b} \Lambda_{\mathbf{R},ab}(\tau) \left[ \Tr [\bar{\Phi}_{\mathbf{R}}(\tau)F^\dagger_a  F_b \Phi_\mathbf{R}(\tau)] - \bar{f}_{\mathbf{R}a}(\tau) f_{\mathbf{R}b}(\tau) \right]. \label{eq:L_gRISB}
\end{align}

Before making the mean-field decoupling between the auxiliary-fermion and slave-boson sectors explicit, we promote the composite quantities $R_{\mathbf{R},a\alpha} \big[ \bar{\Phi}_{\mathbf{R}}, \Phi_{\mathbf{R}} \big]$ and $\Tr [\bar{\Phi}_{\mathbf{R}}F^\dagger_a  F_b \Phi_\mathbf{R}]$ to independent complex matrix fields $R_{\mathbf{R},a\alpha} \big[ \bar{\Phi}_{\mathbf{R}}, \Phi_{\mathbf{R}} \big] \rightarrow R_{\mathbf{R},a\alpha} (\tau)$ and $\Tr [\bar{\Phi}_{\mathbf{R}}F^\dagger_a  F_b \Phi_\mathbf{R}] \rightarrow \Delta_{\mathbf{R},ab} (\tau)$.

Their equivalence to the corresponding slave-boson composite fields is enforced by introducing complex Lagrange-multiplier fields:
\begin{equation}
D_{\mathbf{R},a\alpha}(\tau)
\end{equation}
enforcing:

\begin{equation}
\sum_c R_{\mathbf{R},c\alpha}  [\sqrt{\Delta_\mathbf{R}(1-\Delta_\mathbf{R})} ]_{ca} = \Tr [\bar{\Phi}_{\mathbf{R}}F^\dagger_\alpha \Phi_\mathbf{R} C^\dagger_a ]
\end{equation}
and
\begin{equation}
\Lambda^c_{\mathbf{R},ab}(\tau)
\end{equation}
enforcing:
\begin{equation}
\Delta_{\mathbf{R},ab} = \Tr [\bar{\Phi}_{\mathbf{R}}F^\dagger_a  F_b \Phi_\mathbf{R}]
\end{equation}

The new Lagrangian can be written as:

\begin{align}
    \mathcal{L}^{gRISB} =  \mathcal{L}^{gRISB-f} + \sum_{\mathbf{R}}  \big[  \mathcal{L}^{gRISB-\Phi}_{\mathbf{R}} + \mathcal{L}^{gRISB-mix}_{\mathbf{R}} \big]
    \label{eq:L_gRISB_mf}
\end{align}

where:

\begin{align}
    \mathcal{L}^{gRISB-f} = & \sum_{\mathbf{R},a} \bar{f}_{\mathbf{R}a}(\tau) \partial_{\tau} f_{\mathbf{R}a}(\tau) \nonumber \\
    &- \sum_{\mathbf{R},\mathbf{R}'} \sum_{\alpha,\beta} \sum_{a,b} t_{\mathbf{R}\mathbf{R}'}^{\alpha\beta} \, \bar{f}_{\mathbf{R}a}(\tau) {R}_{\mathbf{R},a\alpha}(\tau) {R}_{\mathbf{R}',b\beta}^{*}(\tau) f_{\mathbf{R}'b}(\tau) \nonumber \\
    &- \sum_{\mathbf{R}} \sum_{a,b} \Lambda_{\mathbf{R},ab}(\tau) \bar{f}_{\mathbf{R}a}(\tau) f_{\mathbf{R}b}(\tau) \label{eq:L_gRISB_f_independent} \\
    \mathcal{L}_{\mathbf{R}}^{gRISB-\Phi} = & \sum_{\Omega,n} \bar{\Phi}_{\mathbf{R},\Omega n}(\tau) \partial_{\tau} \Phi_{\mathbf{R},\Omega n}(\tau) \nonumber \\
    &+ \sum_{\Omega,\Omega'} \left[ H_{\mathbf{R}}^{\mathrm{loc}} \right]_{\Omega\Omega'} \sum_n \bar{\Phi}_{\mathbf{R},\Omega n}(\tau) \Phi_{\mathbf{R},\Omega'n}(\tau) \nonumber \\
    &+ E_{\mathbf{R}}(\tau) \left[ \sum_{\Omega,n} \bar{\Phi}_{\mathbf{R},\Omega n}(\tau) \Phi_{\mathbf{R},\Omega n}(\tau) - 1 \right] \nonumber \\
    &+ \sum_{a,b} \Lambda^c_{\mathbf{R},ab}(\tau) \Tr [\bar{\Phi}_{\mathbf{R}}(\tau) F^\dagger_a  F_b \Phi_\mathbf{R} (\tau) ] \nonumber \\
    &+ \sum_{a,\alpha} \left\{ D_{\mathbf{R},a\alpha}(\tau) \Tr [\bar{\Phi}_{\mathbf{R}} (\tau)F^\dagger_\alpha \Phi_\mathbf{R} (\tau) C^\dagger_a ] + \mathrm{c.c.} \right\} \label{eq:L_gRISB_Phi_independent} \\
    \mathcal{L}^{gRISB-mix}_{\mathbf{R}} = &  \sum_{\mathbf{R}} \sum_{a,b} \bigl[ \Lambda_{\mathbf{R},ab}(\tau) + \Lambda^c_{\mathbf{R},ab}(\tau) \bigr] \Delta_{\mathbf{R},ab}(\tau) \nonumber \\
    &+ \sum_{\mathbf{R}} \sum_{a,\alpha,c} \left[ (\sqrt{\Delta_\mathbf{R}(\tau)(1-\Delta_{\mathbf{R}}(\tau))})_{ca} D_{\mathbf{R},a\alpha}(\tau) R_{\mathbf{R},c \alpha}(\tau) + \mathrm{c.c.} \right]. \label{eq:L_gRISB_constraints_independent}
\end{align}

We then simplify the problem by imposing the static saddle point conditions on the matrix fields $\Delta_{\mathbf{R}}, \Lambda_{\mathbf{R}} , R_\mathbf{R},\Lambda^c_{\mathbf{R}}, D_\mathbf{R}, \Phi_\mathbf{R}$, and $E_\mathbf{R}$. This way we recover the well-known ghost-GA/ghost-RISB equations~\cite{Lanata2017bloch,Lanata2022operatorial}. Imposing the saddle point on these fields means treating them as complex classical fields, and the static saddle point equations read:

\begin{align}
    \frac{\delta \mathcal{L}}{\delta \Lambda_{\mathbf{R},ab}}=0 \rightarrow & \langle f^\dagger_{\mathbf{R}a} f_{\mathbf{R}b}\rangle= \Delta_{\mathbf{R},ab} \\
    \frac{\delta \mathcal{L}}{\delta \Lambda^c_{\mathbf{R},ab}}=0 \rightarrow & \Tr [\bar{\Phi}_{\mathbf{R}}F^\dagger_a  F_b \Phi_\mathbf{R}]= \Delta_{\mathbf{R},ab} \\
    \frac{\delta \mathcal{L}}{\delta R_{\mathbf{R},a \alpha}}=0 \rightarrow & \sum_{\mathbf{R}^\prime \neq \mathbf{R}} \sum_{b,\beta}t^{\alpha \beta}_{ab} R^*_{\mathbf{R}^\prime, b \beta} \langle f^\dagger_{\mathbf{R}a} f_{\mathbf{R}^\prime b} \rangle = \sum_{c} [\sqrt{\Delta_\mathbf{R}(1-\Delta_{\mathbf{R}})}]_{ac} D_{\mathbf{R},c\alpha} \\
    \frac{\delta \mathcal{L}}{\delta D_{\mathbf{R},c \alpha}}=0 \rightarrow &  \sum_c R_{\mathbf{R},c\alpha}  [\sqrt{\Delta_\mathbf{R}(1-\Delta_\mathbf{R})} ]_{ca} = \Tr [\bar{\Phi}_{\mathbf{R}}F^\dagger_\alpha \Phi_\mathbf{R} C^\dagger_a ]\\
    \frac{\delta \mathcal{L}}{\delta \Delta_{\mathbf{R},mn}}=0 \rightarrow &  [\Lambda^c_{\mathbf{R}}+\Lambda_{\mathbf{R}}]_{mn}  = \Big[ 
 \sum_{\alpha c a} \mathcal{R}_{\mathbf{R}, c \alpha} \frac{\delta [\Delta_\mathbf{R}(1-\Delta_\mathbf{R})]^{-1/2}_{ca} }{\delta 
 [\Delta_{\mathbf{R}}]_{mn} }
 D_{\mathbf{R},a \alpha} + c.c.  \Big] \label{eq:saddle_delta} \\
 \frac{\delta \mathcal{L}}{\delta E_{\mathbf{R}}}=0 \rightarrow & \sum_{\Omega n} \bar{\Phi}_{\mathbf{R},\Omega n} \Phi_{\mathbf{R},\Omega n}= 1 \\
 \frac{\delta \mathcal{L}}{\delta \bar{\Phi}_{\mathbf{R},\Omega n}}=0 \rightarrow & \sum_{\Omega^\prime n^\prime}[H_{\mathbf{R}}^{emb}]_{ \Omega n , \Omega^\prime n^\prime } \Phi_{\mathbf{R},\Omega^\prime n^\prime}= E_\mathbf{R} \Phi_{\Omega n} 
\end{align}

Where in the last equation we made use of the embedding mapping~\cite{lanata2015,Lanata2017bloch} where the bosonic field saddle point equation can be seen as an eigenvalue problem for a impurity problem wavefunction whose weight on the computational basis are related to the slave-bosons amplitudes as

\begin{equation}
    |\Psi_{emb}\rangle = \sum_{\Omega n}   \Phi_{\Omega n}  s(n) | \Omega_\mathbf{R} \rangle \otimes [\hat{U}_{PH} |n_\mathbf{R} \rangle ] 
\end{equation}
where $s(n) = e^{i(\pi/2)N(n)[N(n)-1]}$ and $U_{PH}$ is a particle-hole transformation acting on the Hilbert space spanned by bath states $|n\rangle $ and $N(n)$ is the number of electrons in such state.

These equations are equivalent to the ghost-Gutzwiller Approximation~\cite{Lanata2017bloch} and ghost-RISB~\cite{Lanata2022operatorial} with the simple addition of phononic fields in the impurity Hamiltonian.

\section{Proof that the Holstein bipolaronic overlap is an upper bound of the s-wave superconductive order parameter} \label{app:orthogonality}

The ground state of the embedding Hamiltonian can be written generically as:
\begin{equation}
    | \Psi \rangle = \sum_{m,\Gamma,\Upsilon} \psi_{m,\Gamma,\Upsilon} |m , \Gamma,\Upsilon\rangle
\end{equation}

where $|m\rangle$ is a basis for the bath Fock space, $|\Gamma\rangle$ is a basis of the impurity Fock space and $|\Upsilon\rangle$ is a basis for the phononic Fock space.
We define the charge-resolved phonon distributions for the doublon and holon sectors as
\begin{align}
P_{2}(X)
=&  \sum_{m}  \langle m,\uparrow  \downarrow ,X|\Psi \rangle  \langle \Psi| m,\uparrow  \downarrow , X\rangle \nonumber \\
=&
\sum_m
\left|
\sum_\Upsilon \phi_\Upsilon (X)\psi_{m,\uparrow\downarrow,\Upsilon}
\right|^2
\end{align}
and
\begin{align}
P_{0}(X)
=&  \sum_{b}  \langle m,0,X|\Psi \rangle  \langle \Psi| m,0, X\rangle \nonumber \\
=&
\sum_m
\left|
\sum_\Upsilon \phi_\Upsilon(X)\psi_{m,0,\Upsilon}
\right|^2
\end{align}
Where $|X\rangle$ is the eigenstate of the phononic displacement operator with eigenvalue $X$.
The $s$-wave superconductive order parameter can be computed as:
\begin{align}
    |\langle \Psi| c^\dagger_\uparrow  c^\dagger_\downarrow | \Psi \rangle| = \Big| \sum_{\Gamma,m,\Upsilon} \sum_{\Gamma^\prime , m^\prime , \Upsilon^\prime}\Big[ &  \langle m^\prime,\Gamma^\prime,\Upsilon^\prime |  c^\dagger_\uparrow  c^\dagger_\downarrow|  m,\Gamma,\Upsilon\rangle  \psi^\star_{m^\prime,\Gamma^\prime,\Upsilon^\prime}  \psi_{m,\Gamma,\Upsilon} \Big] \Big|
\end{align}

Clearly only the contribution coming from $|\Gamma\rangle=|0\rangle$ and $\langle \Gamma^\prime | = \langle \uparrow \downarrow |$ are non zero. Adding the phononic identity $\hat{I}=\int dX |X\rangle \langle X |$ in the equation leads to:
\begin{equation}
    \Big| \langle \Psi| c^\dagger_\uparrow  c^\dagger_\downarrow | \Psi \rangle \Big| = \Big| \sum_{m}     \int dX [ \sum_{\Upsilon^\prime} \phi_{\Upsilon^\prime}^\star(X) \psi^\star_{m,\uparrow\downarrow,\Upsilon^\prime}] [  \sum_\Upsilon \phi_{\Upsilon}(X) \psi_{m,0,\Upsilon}] \Big|
\end{equation}
Where $\phi_\Upsilon(X)=\langle X | \Upsilon \rangle$ is the eigenstate of the bosonic quantum harmonic oscillator with quantum number $\Upsilon $.
Making use of some standard inequalities:
\begin{align}
    \Big| \langle \Psi| c^\dagger_\uparrow  c^\dagger_\downarrow | \Psi \rangle \Big| =&  \Big|\sum_{m}     \int dX [ \sum_{\Upsilon^\prime} \phi_{\Upsilon^\prime}^\star(X) \psi^\star_{m,\uparrow\downarrow,\Upsilon^\prime}] [  \sum_\Upsilon \phi_{\Upsilon}(X) \psi_{m,0,\Upsilon}] \Big|\\
    \leq& \sum_{m}     \int dX | \sum_{\Upsilon^\prime} \phi_{\Upsilon^\prime}^\star(X) \psi^\star_{m,\uparrow\downarrow,\Upsilon^\prime}| \cdot| \sum_\Upsilon \phi_{\Upsilon}(X) \psi_{m,0,\Upsilon}| \\
    = & \sum_{m}     \int dX \sqrt{ |\sum_{\Upsilon^\prime} \phi_{\Upsilon^\prime}^\star(X) \psi^\star_{m,\uparrow\downarrow,\Upsilon^\prime}|^2 } \cdot \sqrt{ |\sum_\Upsilon \phi_{\Upsilon}(X) \psi_{m,0,\Upsilon}|^2 } \\
    \leq &   \int dX \Big( \sqrt{ \sum_{m}|\sum_{\Upsilon^\prime} \phi_{\Upsilon^\prime}^\star(X) \psi^\star_{m,\uparrow\downarrow,\Upsilon^\prime}|^2 } \Big) \Big(  \sqrt{ \sum_{m}|\sum_\Upsilon \phi_{\Upsilon}(X) \psi_{m,0,\Upsilon}|^2 } \Big) \\
    = &   \int dX \sqrt{P_{2}(X)} \sqrt{P_{0}(X)} = \mathcal{I}_{dh}
\end{align}

Therefore, the overlap between the square root of the charge-resolved distribution function of the phononic field is an upper bound for the pairing amplitude.

\end{widetext}

\end{document}